\documentclass[fleqn,twoside]{article}
\usepackage{espcrc2}

\usepackage{graphicx}

\newcommand{\AmS}{{\protect\the\textfont2
  A\kern-.1667em\lower.5ex\hbox{M}\kern-.125emS}}

\title{Magnetoresistance and magnetic anisotropy in
La$_{0.5}$Sr$_{0.5}$CoO$_{3-\delta}$ film}

\author{B. I. Belevtsev\address[ILTPE]{B. Verkin Institute for Low 
Temperature Physics \& Engineering, Kharkov, 61103, Ukraine} 
\thanks{Corresponding author.   fax ++380-572-335593;
e-mail:  belevtsev@ilt.kharkov.ua},
V. B. Krasovitsky\addressmark[ILTPE],
A. S. Panfilov\addressmark[ILTPE],
I. N. Chukanova\address[ISS]{Institute for Single Crystals, 61001,
Kharkov, Ukraine}}

\begin{document}

\begin{abstract}
The magnetic and transport properties of La$_{0.5}$Sr$_{0.5}$CoO$_{3-\delta}$
film grown on a LaAlO$_3$ substrate by pulsed-laser deposition are 
studied. The properties are found to be influenced by a combined 
influence of the magnetic anisotropy and inhomogeneity. Magnetoresistance 
anisotropy is determined by the shape anisotropy and the strain-induced 
magnetic anisotropy due to the film-substrate lattice interaction.   
Indications of the temperature-driven spin reorientation transition from
an out-of plane orderded state at low temperatures to an in-plane 
ordered state at high temperatures as a result 
of competition between the mentioned anisotropy sources are found.
\vspace{1pc}
\end{abstract}

\maketitle

\section{INTRODUCTION} 
\label{int}
Hole-doped lanthanum cobaltates of the type  La$_{1-x}$Sr$_{x}$CoO$_{3}$ have
attracted much attention in recent years due to their unique magnetic 
and transport properties \cite{itoh,gooden}. Study of this system is also
important for understanding the nature of colossal magnetoresistance 
in the related oxides, doped manganites \cite{dagotto}. 
For technical application, the epitaxial films of these compounds 
are mainly implied to be used. In that case the shape anisotropy and the 
film-substrate lattice interaction can induce magnetization anisotropy
and, therefore, magnetoresistance (MR) anisotropy (bulk samples of these 
compounds show no marked magnetic and MR anisotropy). This point 
was  studied rather intensively  in manganite films (see \cite{belev1} and 
references therein), but it is hardly to find in literature some studies 
of this type  for doped cobaltates. Beside this, the properties of 
doped cobaltates  are influenced by their (extrinsic and intrinsic) 
magnetic inhomogeneity. In this report we 
present a study of La$_{0.5}$Sr$_{0.5}$CoO$_{3-\delta}$ film which 
demonstrates a combined influence of the magnetic anisotropy and 
inhomogeneity on its transport and magnetic properties.

\section{EXPERIMENTAL}
The La$_{0.5}$Sr$_{0.5}$CoO$_{3-\delta}$ film (about 220 nm thick) 
was grown by pulsed-laser deposition (PLD) on a (001) oriented LaAlO$_3$ 
substrate. The ceramic target used was prepared by a standard solid-state
reaction technique. A PLD system with an Nd-YAG laser operating at 1.06 
$\mu$m was used to ablate the target. The pulse energy was about 0.39 J with 
a repetition rate  of 12 Hz and pulse duration of 10 ns.
The film was deposited with a substrate temperature of 880$\pm 5^\circ$C
in oxygen atmosphere at a pressure of about 8~Pa. The film was cooled 
down to room temperature after deposition at an oxygen pressure about 
$10^5$~Pa. The target and film were characterized by X-ray diffraction 
(XRD) study. 
\par
The film resistance, as a function of temperature and magnetic field
$H$ (up to 20 kOe), was measured using a standard four-point technique. 
The field was applied parallel or perpendicular to
the film plane. In both cases it was perpendicular to the transport 
current. The magnetization, $M$, was measured in a Faraday-type 
magnetometer.

\section{RESULTS AND DISCUSSION}

\begin{figure}[htb]
\vspace{-20pt}
\centering\includegraphics[width=0.47\textwidth]{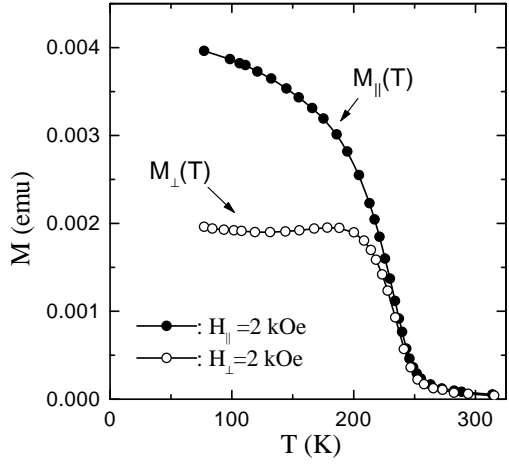}
\vspace{-40pt}
\caption{Temperature dependences of the magnetization of the film studied in
the directions parallel and perpendicular to the film plane. 
}
\label{m(t)}
\end{figure}

Temperature dependences of the film magnetization for the field directions 
parallel [$M_{\parallel}(T)$] and perpendicular [$M_{\perp}(T)$] to the 
film plane are shown in Fig.~1. The Curie temperature, $T_c$, is found to 
be about 250~K. The $M_{\parallel}(T)$ behaviour is quite common for 
ferromagnetic (FM) metals. At fairly high field used, 2~kOe, the 
$M_{\parallel}(T)$ curve is found to be well above the $M_{\perp}(T)$ 
curve. It is reasonable to suppose that this is determined mainly by the 
shape anisotropy. Closer inspection shows, however, that  $M_{\perp}(T)$ 
behavior cannot be attributed solely to the shape-anisotropy effect: 
$M_{\perp}(T)$ and $M_{\parallel}(T)$ are practically equal in rather 
broad temperature range below $T_c$, then (going to lower temperature) 
the $M_{\perp}(T)$ curve goes rather abruptly well below the 
$M_{\parallel}(T)$ curve and becomes non-monotonic with a pronounced 
increase in $M_{\perp}(T)$ at low temperatures. These $M_{\perp}(T)$ 
features can be caused by the strain-induced magnetic anisotropy due 
to the film-substrate interaction. This guess is 
supported by the XRD study which has revealed that the film has 
an out-of-plane tensile strain. For materials with the positive
magnetostriction this must favours an out-of-plane easy magnetization. 

\begin{figure}[htb]
\vspace{-15pt}
\centering\includegraphics[width=0.47\textwidth ]{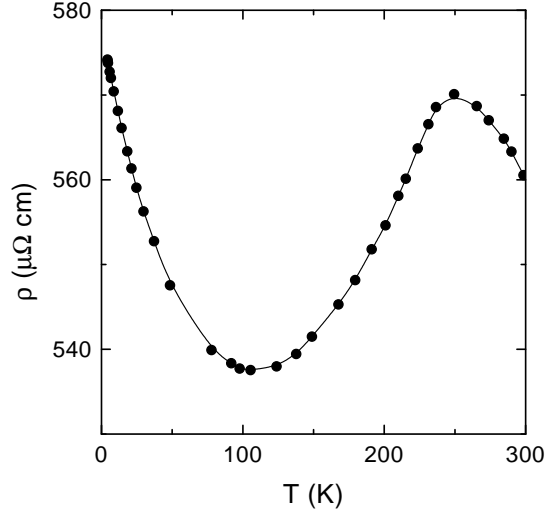}
\vspace{-30pt}
\caption{Temperature dependence of the film resistivity. 
}
\label{r(t)}
\end{figure}

The temperature dependence of the resistivity, $\rho (T)$, is 
non-monotonic (Fig.~2) with a maximum at $T\approx 250$~K and a minimum 
at $T\approx 107$~K.  
La$_{0.5}$Sr$_{0.5}$CoO$_{3-\delta}$ samples with fairly perfect 
crystalline structure and $\delta$ close to zero are  
known to be metallic ($d\rho/dT > 0$) in the whole range below and above 
$T_c$ \cite{gooden}. The $\rho (T)$ behaviour in Fig.~2 reflects 
inhomogeneous structure of the film and some oxygen deficiency. Due to the 
last factor, the hole 
concentration is less then a nominal one (at $\delta=0$). 
This is responsible
for a resistance peak at $T=250$~K which is common for low-doped samples 
with $0.2 \geq x \geq 0.3$ \cite{gooden}. The low temperature 
resistance minimum  is typical for systems of FM regions (grains or 
clusters) with rather weak interconnections. The 
inhomogenenous structure can be determined by technological factors
of sample preparation (causing the polycrystalline structure with rather
high tunneling bariers between the grains) or by the phase separation into
the hole-rich and hole-poor phase \cite{itoh,gooden}. For an extended
discussion of these points for cobaltate films see reference 
\cite{belev2}. 

\begin{figure}[htb]
\vspace{-20pt}
\centering\includegraphics[width=75mm]{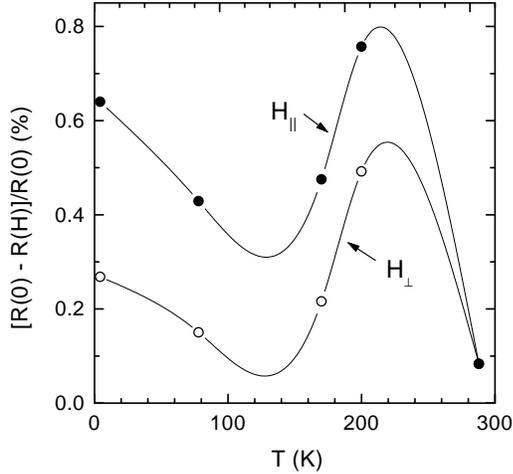}
\vspace{-30pt}
\caption{Magnetoresistance at $H=20$~kOe for fields parallel and
perpendicular to the film plane.  
}
\label{r(h)}
\end{figure}
\par
The MR is found to be anisotropic. The absolute values of negative MR
in fields parallel to the film plane are considerably above those in
perpendicular fields (Fig.~3). This MR anisotropy takes place only 
in FM state and disappears for $T>T_c$ (Fig.~3). An increase in MR with
decreasing temperature (in the range well below $T_c$) is one more
indication of poor enough connectivity between the FM grains (or clusters)
in the film. 
\par
The data presented in Fig.~3 are pertaining to negative MR for fairly
high fields. In general, the MR curves are hysteretic and have 
specific features in low-field range (Fig.~4). Actually, their behaviour 
correlates with that of magnetization curves. In particular, the field
$H=H_p$, at which MR has a peak (Fig.~4), corresponds to value of the 
coercive force ($H_c$). The value of $H_p$ decreases with increasing 
temperature 
and goes to zero with approaching $T_c$. The magnitude of positive MR in 
the low-field range, $\Delta R(H_{p}) = [R(H_{p})-R(0)]/R(0)$, is some 
measure of the remanent magnetization. 

\begin{figure}[htb]
\vspace{-20pt}
\centering\includegraphics[width=0.47\textwidth]{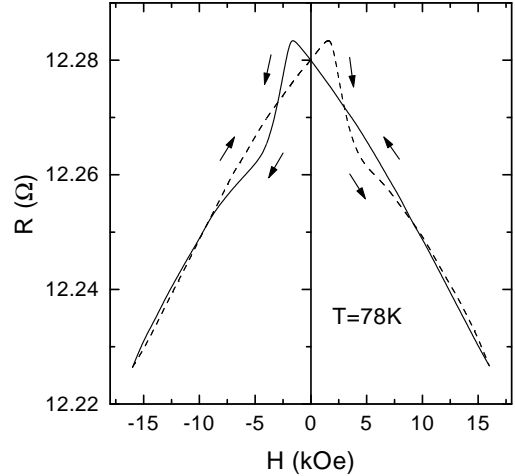}
\vspace{-40pt}
\caption{Magnetoresistive hysteresis. Field is parallel to the 
film plane and perpendicular to the transport current. 
}
\label{hister}
\end{figure}

\par
We found that $H_p$
and $\Delta R(H_{p})$ depend on the field direction and reflect in
this way the magnetization anisotropy. In particular, at $T\simeq 4.2$~K
the value of $H_p$ in the out-of-plane field is 
less than that in the in-plane field, and $\Delta R(H_{p})$ value is 
higher for the out-of-plane field direction. The opposite relations
(that is, lesser $H_p$ values and higher $\Delta R(H_{p})$ values for
the in-plane field) are found for higher temperatures $T \geq 70$~K.
This implies that at low temperatures the out-of-plane magnetization is
favoured, whereas at higher temperatures the in-plane magnetization
becomes dominant. The pronounced increase in $M_{\perp}(T)$ at low 
temperatures (Fig.~1) supports additionally this suggestion. All these 
are doubtless indications of the temperature-driven spin reorientation 
transition which is a result of competition between the shape anisotropy
and the strain-induced anisotropy. 
This transition has been studied intensively for films of common FM 
metals (see, e.~g. \cite{hu}), but was never mentioned for manganite 
or cobaltate films.

\end{document}